\documentclass[authoryear]{elsarticle}

\usepackage{hyperref,csvsimple,booktabs,longtable,lscape,graphicx}
\usepackage{amsmath}
\usepackage{booktabs}
\usepackage{float}
\usepackage{natbib}
\usepackage[T1]{fontenc}
\usepackage[icelandic,english]{babel}
\usepackage[utf8]{inputenc}

\makeatletter
\def\ps@pprintTitle{%
	\let\@oddhead\@empty
	\let\@evenhead\@empty
	\def\@oddfoot{}%
	\let\@evenfoot\@oddfoot}
\makeatother

\begin{document}

\begin{frontmatter}

\title{An overview of the marine food web in Icelandic waters using Ecopath with Ecosim}

\author[mymainaddress]{Joana P.C. Ribeiro\corref{mycorrespondingauthor}}

\cortext[mycorrespondingauthor]{Corresponding author}

\ead{pcrjoana@gmail.com}

\author[mysecondaryaddress]{Bjarki Þ. Elvarsson}

\author[mymainaddress]{Erla Sturludóttir}

\author[mymainaddress]{Gunnar Stefánsson}

\address[mymainaddress]{Science Institute, University of Iceland, Dunhagi 7, 107 Reykjavík, Iceland}

\address[mysecondaryaddress]{Marine and Freshwater Research Institute, Skúlagata 4, 101 Reykjavík, Iceland}

\begin{abstract}
Fishing activities have broad impacts that affect, although not exclusively, the targeted stocks. 
These impacts affect predators and prey of the harvested species, as well as the whole ecosystem it inhabits. 
Ecosystem models can be used to study the interactions that occur within a system, including those between different organisms and those between fisheries and targeted species. 
Trophic web models like Ecopath with Ecosim (EwE) can handle fishing fleets as a top predator, with top-down impact on harvested organisms.
The aim of this study was to better understand the Icelandic marine ecosystem and the interactions within. 
This was done by constructing an EwE model of Icelandic waters. 
The model was run from 1984 to 2013 and was fitted to time series of biomass estimates, landings data and mean annual temperature. 
The final model was chosen by selecting the model with the lowest Akaike information criterion.
A skill assessment was performed using the Pearson’s correlation coefficient, the coefficient of determination, the modelling efficiency and the reliability index to evaluate the model performance.
The model performed satisfactorily when simulating previously estimated biomass and known landings. Most of the groups with time series were estimated to have top-down control over their prey. These are harvested species with direct and/or indirect links to lower trophic levels and future fishing policies should take this into account. This model could be used as a tool to investigate how such policies could impact the marine ecosystem in Icelandic waters.\par
\end{abstract}


\end{frontmatter}

\section{Introduction}
Fisheries affect targeted stocks, as well as the whole ecosystem. Fisheries can have negative effects due to maximizing stock profitability. 
These include damage to the habitat and impacts to predators and prey of the fished species \citep{RN185}. 
The need for fisheries management arose from overexploitation of fish stocks. 
Since they were first developed, different measures of fisheries management have been employed across systems and their effectiveness has been previously studied \citep{RN188}. Fishing policies, or lack thereof, have not always prevented overexploitation \citep{RN214}.\par
Fisheries management strategies have had a tendency to focus on a single species and not on wider, more realistic multi species scenarios \citep{RN218,RN185}.
Ecosystems harbour different species within them, which interact with each other and with their environment. 
The species in a given system are connected through trophic links and other types of ecological interactions. 
Ecosystem models can be used to gain better understanding of the underlying dynamics within a system. 
Trophic links in particular can be studied using food web models, also referred to as trophic models. 
When a system is harvested, the interaction between the targeted species and the fishing fleet is introduced in the system. 
In this case, fishing fleets can be handled by food web models as a top predator which interact with targeted stocks through top-down trophic control.\par
A widely used food web model is Ecopath with Ecosim \citep[EwE;][]{RN17,RN93,RN69,RN176}. 
Typical uses of EwE include answering ecological questions, quantifying trophic flows and studying the food-web structure and investigating potential impacts of fisheries on ecosystems. 
EwE has also been used as a fisheries management scenarios evaluation tool \citep[for a complete overview on EwE capacities and limitations see][]{RN18}.\par
Two EwE models for Icelandic waters had previously been published \citep[Buchary, E.A.; Mendy, A.N. and Buchary, E.A., in][]{RN208,RN171}. The earlier model \citep[Buchary, E.A. and Mendy, A.N. in ][]{RN208, RN171} modelled the trophic web in Icelandic waters in the year 1997 and did not include Ecosim. The latter model \citep[Buchary E.A. in][]{RN208} was a reconstruction of the food web in Icelandic waters in 1950 and included a dynamic component that simulated the trophic dynamics between 1950 and 1997. These models did not cover more recent dynamics nor were they evaluated using a skill assessment.
Other work on trophic interactions in Icelandic waters include specific predator-prey interactions \citep{RN178}, isotope analyses of organisms in the Icelandic sea \citep{RN179} and end to end modelling of the marine ecosystem \citep{RN220}.\par 
The marine environment around Iceland has been exploited for centuries. In the 20\textsuperscript{th} century, fisheries were the main source of income in Iceland \citep{RN142}. Economically important species include demersal fish such as cod (\textit{Gadus morhua}), saithe (\textit{Pollachius virens}), haddock (\textit{Melanogrammus aeglefinus}), redfish (\textit{Sebastes spp.}) and Greenland halibut (\textit{Rinhardtius hippoglossoides}), as well as pelagic species such as herring (\textit{Clupea harengus}), capelin (\textit{Mallotus villosus}) and, in later years, blue whiting \citep[\textit{Micromesistius poutassou};][]{RN2}.\par
The aim of this study was to gain better understanding of the ecosystem in Icelandic waters by using EwE to model the trophic interactions between species in the ecosystem, as well as the interactions between fisheries and targeted species. The aim was also to evaluate the performance of the model with a skill assessment.  \par

\section{Methods}

\subsection{Study area} 
Iceland is an island located in the North Atlantic ocean, at the junction of the Middle Atlantic Ruidge and the Greenland-Scotland ridge \citep{RN2}. Its exclusive economic zone (EEZ) spans an area of 758 000 km$^2$ \citep{RN142} (\figurename{ 1}). This area includes the continental shelf around Iceland.\par

\begin{figure}[H]
	\includegraphics[scale=0.5]{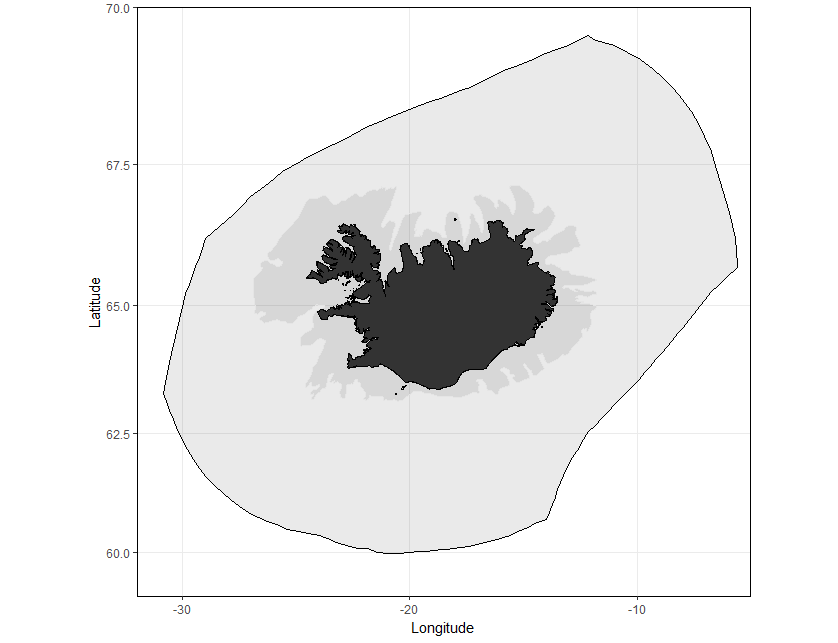}\par
	\caption{Icelandic EEZ with Iceland marked in black and highlighted continental shelf.}
	\normalsize
\end{figure}

The Icelandic waters harbour organisms ranging from phytoplankton to marine mammals, as well as sea birds. 
Some of these are differently distributed along the ecosystem in Icelandic waters, due to different water salinity and temperature between the southern and western areas and the northern and eastern parts \citep{RN2}. 
The southern and western areas have warm and saline Atlantic water, while the water in northern and eastern areas is a mix of Atlantic, Arctic and Polar water \citep{RN217, RN216, RN2}.
The different habitats resulting from this difference in water composition is exploited by some species, including commercially important fish stocks. These spawn in southern and western waters and feed off northern and western waters \citep{RN215,RN2}.   
\par

\subsection{Mass-balance model}
To study the marine ecosystem of Icelandic waters, EwE \citep{RN17,RN93,RN69,RN176} was chosen as a tool to give a simple overview of the entire system. EwE is composed of two modelling stages, Ecopath and Ecosim. Ecopath is the mass-balance part of EwE and its master equation is as follows:\par
\begin{equation}%
	B_i\left(\frac{P}{B}\right)_i  = \sum_{j}B_j\left(\frac{Q}{B}\right)_j DC_{ij} + Y_i + E_i + BA_i + B_i\left(\frac{P}{B}\right)_i(1-EE_i)
\end{equation}
\bigskip

Here, $B_i$ is the biomass of functional group $i$, $(P/B)_i$ is the production to biomass ratio, equivalent to total mortality ($Z$) in closed systems and corresponding to the sum of natural mortality ($M$) and fishing mortality ($F$), $B_j$ is the biomass of predator $j$, $(Q/B)_j$ is the consumption to biomass ratio of $j$, $DC_{ij}$ is the proportion of $i$ found in the stomach of $j$, $Y_i$ is the yield, $E_i$ is the net migration, $BA_i$ is the biomass accumulation and $EE_i$ is the ecotrophic efficiency, which corresponds to the production of $i$ explained by the model. $F$ is calculated in Ecopath as the ratio $Y/B$. \par

\subsubsection{Functional groups}
The Icelandic Ecopath model was comprised of 35 functional groups, of which six contained two stanzas each.
The most relevant species of phytoplankton, zooplankton, invertebrates, fish, marine mammals and sea birds were included in the 35 functional groups used in the EwE model here presented (\tablename{ 1}). Species were chosen as multi-stanza and/or individual functional groups depending on their economical importance and/or their role on the ecosystem and available biomass estimates and/or landings data. The aggregation of different species in a single functional group was based on species type and habitat use (\tablename{ A.1).\par
Commercial species were targeted by a single fishing fleet, without discrimination regarding fishing gear.\par

\begin{table}[H]
	\begin{center}
		\caption{List of the functional groups for the Icelandic waters EwE model.}\par
		\footnotesize
		\begin{filecontents*}{FG.csv}
,Functional groups,
Seabirds,Herring* (0-3),Commercial demersal
Minke whale,Herring* (4+),Other demersal
Baleen whales,Redfish* (0-7),Cephalopod
Tooth whales,Redfish* (8+),Mollusc
Pinniped,Greenland halibut* (0-5),Lobster
Greenland shark,Greenland halibut* (6+),Shrimp
Dogshark,Capelin,Benthos
Skate rays,Blue whiting,Jellyfish
Cod* (0-3),Mackerel,Krill
Cod* (4+),Sandeel,Crustacean zooplankton
Saithe* (0-3),Large pelagic,Small zooplankton
Saithe* (4+),Small pelagic,Phytoplankton
Haddock* (0-2),Flatfish,Detritus
Haddock* (3+),Other codfish,
,,
*Multi-stanza group,,
\end{filecontents*}
\csvautobooktabular{FG.csv}
		\normalsize	
	\end{center}
\end{table}

\subsubsection{Initial parameters} 
Ecopath balances the model by solving a series of linear equations (Eq. 1). For each equation there must be an unknown value. The four basic values in Ecopath are $B$, $P/B$, $Q/B$ and $EE$. Ideally, $B$, $P/B$ and $Q/B$ should be input values. However, if estimates of $B$ are lacking, the $EE$ can be used as an input value to estimate $B$.
In multi-stanza groups, $B$, $P/B$ and $Q/B$ have to be inserted for the leading stanza, whereas only $P/B$ needs to be inserted for all stanzas. 
Even though $P/B$ and $Q/B$ can be estimated, it is highly recommended that both are part of the input. 
The assumption of a closed system was made when modelling the marine ecosystem in Iceland. In a closed system, there is no $BA$ nor $E$, thus these values were set to 0.\par
Ecopath was set up for the year 1984 and the input biomass and landings corresponded to estimates for this year or for the next year with available data. A comprehensive list of the literature used can be found in \tablename{ 2}.\par

\begin{table}[H]
	\caption{Literature sources of biomass and $Q/B$ estimates and of landings.}\par
	\footnotesize
	\begin{filecontents*}{Lit.csv}
Functional group,Literature used
Seabirds,\citet{RN52}
,\citet{RN13}
,\citet{RN161}
Minke whale;,\citet{RN76}  
Baleen whales and,\citet{RN207}
Tooth whales,\citet{RN205}
,\citet{RN206}
,\citet{RN148}
Pinniped,\citet{RN103}
,\citet{RN13}
All fish groups,\citet{RN148} 
,\citet{RN152}
Greenland shark (\textit{Somniosus microcephalus}; biomass),\citet{RN56}
Cod ($Q/B$),\citet{RN102}
"Cephalopod; lobster; shrimp,  ;",\citet{RN13}
Mollusc; jellyfish; benthos;,
Small zooplankton and crustacean zooplankton,
Phytoplankton, \citet{RN151}
\end{filecontents*}
\csvautobooktabular{Lit.csv}\par
\end{table}

No biomass was found for functional groups dogshark, skate rays, sandeel, large pelagic, small pelagic, flatfish, other demersal, lobster, shrimp, benthos, jellyfish and both zooplankton groups. Since these were not harvested in Iceland in 1984, it was not possible to estimate biomass and as such the $EE$ was used instead. The $EE$ of dogshark and skate rays was set to 0.500. For all remaining groups, $EE$ was set to 0.950.
The choice of $EE$ values was made based on recommendations in the literature \citep{RN153}.\par

\subsubsection{Diet composition} 
Stomach content data of most fish groups was provided by the MFRI \citep{RN148} and then transformed into diet composition using R \citep{RN150}, to be later used in Ecopath. Diet composition of all other groups, as well as of small pelagic fish, other demersal fish and invertebrates was set according to the literature (\tablename{ 3}).\par

\begin{table}[H]
 \begin{center}
	\caption{Literature sources of diet composition.}\par
	\footnotesize
	\begin{filecontents*}{Lit2.csv}
Functional group,,,Literature used
Seabirds,,,\citet{RN52}
Minke whale,,,\citet{RN87}
Baleen whales,,,\citet{RN76}
Tooth whales,,,\citet{RN12}
,,,\citet{RN76}
,,,\citet{RN88}
Pinnipeds,,,\citet{RN164}
Large pelagic fish,,,\citet{RN104}
Small pelagic fish,,,\citet{RN86}
Other demersal fish,,,\citet{RN104}
All inverebrate groups,,,\citet{RN103}
\end{filecontents*}
\csvautobooktabular{Lit2.csv}\par
 \end{center}
\end{table}

\subsubsection{Model balancing} 
After initialising Ecopath for the first time, two groups had $EE>1$. These were functional group herring 0-3 ($EE=1.201$) and capelin ($EE = 2.328$). In order to balance the model, $P/B$ of the herring 0-3 group was raised from 0.490 to 0.570 and the $EE$ was set to 0.950 for capelin. The latter decision was based on the hypothesis that capelin biomass has been underestimated in assessments \citep{RN102}. According to a study by \citet{RN102}, assessments of capelin biomass have to be approximately doubled to explain consumption by predators.\par

\subsection{Dynamic model}
Ecosim is the dynamic part of the EwE modelling suite \citep{RN93,RN176,RN18}. Ecopath  estimates are used as initial values in Ecosim, which runs a simulation for a user-defined period. The Ecosim model presented here had a time span of 30 years. Ecosim uses a system of differential equations to give a time series of simulated biomass and landings:\par
\begin{equation}%
   \frac{dB_i}{dt} = g_i\sum_{j}Q_{ji} - \sum_{j}Q_{ij} + I_i - (M0_i + F_i + e_i) B_i
\end{equation}
\bigskip

Here, $dB_i/dt$ represents the rate of change in biomass of functional group $i$, described in terms of its biomass, $B_i$, $g_i$ is net growth efficiency ($P/Q$), $Q_{ji}$ is the consumption of $j$ by $i$, $Q_{ij}$ is the predation by $j$ on $i$, $M0_i$ is the mortality not explained by the model, $F_i$ is the fishing mortality rate, $e_i$ is the emigration rate and $I_i$ is the immigration rate.\par

Consumption in Ecosim is calculated based on the foraging arena theory \citep{RN93, RN1}, in which the biomass of the prey is split into vulnerable and non-vulnerable pools. Consumption is computed as:

\begin{equation}%
    \mbox{\Large\( %
	Q_{ij} = \frac{v_{ij} \cdot a_{ij} \cdot B_i \cdot B_j \cdot T_i \cdot T_j \cdot S_{ij} \cdot M_{ij}/D_j}{v_{ij} + v_{ij} \cdot T_i \cdot M_{ij} + a_{ij} \cdot M_{ij} B_j \cdot S_j \cdot T_j/D_j} %
	\)} %
\end{equation}
\bigskip

Here, $v_{ij}$ is the vulnerability of prey $i$ to predator $j$, $a_{ij}$ is the rate of effective search of $i$ by $j$, $T_i$ is the relative feeding time of $i$, $T_j$ is the relative feeding time of $j$, $S_{ij}$ is the representation of seasonal or long-term forcing effects, $M_{ij}$ represents mediation forcing effects and $D_j$ is the impact of handling time as a limit to consumption.\par

\subsubsection{Model fitting}

Ecosim benefits from reference temporal data used to calibrate the model. This reference data includes biomass and/or landings. In addition, biomass, landings, $F$ and environmental variables can be forced to drive the model. To calibrate the model, a time series of both forced and reference values was used (\tablename{ 4}).\par

\begin{table}[H]
	\begin{center}
		\caption{Description of the time-series used to calibrate Ecosim. F - Fishing mortality (forced); FB - Forced biomass; RB - Reference biomass; RC - Reference landings; EV - Environmental variables.}\par
		\footnotesize
		\begin{filecontents*}{TS-description.csv}
Type,Group,Period
F,Cod 4+; Saithe 4+; Haddock 3+;,1984-2013
,Herring 4+;  Redfish 8+; Capelin;,
,Blue whiting; Flatfish; Other codfish; ,
,Mackerel;,2005-2013
,Commercial demersal; ,1994-2013
FB,Mackerel;,1984-2013
RB,Cod 4+; Saithe 4+; Haddock 3+; ,1984-2013
,Herring 4+; Redfish 8+; Blue whiting;,
RC,Cod 4+; Saithe 4+; Haddock 3+; ,1984-2013
,Herring 4+; Redfish 8+; Greenland halibut 6+; ,
,Capelin;  Flatfish; Blue whiting; ,
,Mackerel;,2005-2013
,Other codfish; Commercial demmersal.,1984-2013
EV,Temperature,1984-2013
\end{filecontents*}
\csvautobooktabular{TS-description.csv}\par
		\normalsize
	\end{center}	
\end{table}

Biomass forcing was used for mackerel. Numbers of mackerel have been increasing in Icelandic waters since around 2007 \citep{RN106}. Initial biomass of mackerel was set to low values (0.200 t.km$^{-2}$), as new functional groups cannot be introduced in Ecosim. The invasion by mackerel was then simulated by forcing its biomass from 2005 on. This was the chosen method to handle the increasing numbers of mackerel in the ecosystem as a method to simulate species invasion in EwE. 
Average annual temperature was used as forcing function for PP.
The time series was provided the MFRI \citep{RN148}.\par
The model was fitted to the time series by conducting anomaly and vulnerability searches. The former reduced the total sum of squares (SSQ) by adjusting the forcing function on PP, while the latter reduced SSQ by adjusting vulnerability parameters. 
In EwE, the anomaly search uses spline regression to search for time series values of annual relative PP that could explain productivity shifts impacting biomass across the ecosystem \citep{RN97}. Anomaly searches using 5, 10, 15 and 20 spline points were conducted and model selection was done using the Akaike method \citep{RN219}. The Akaike method for model selection uses the Akaike information criterion (AIC) as a measure of model how well the model fits to the data considering the number of parameters and the lower the AIC, the better the model \citep{RN219}.\par

\subsubsection{Skill assessment}
A skill assessment of the model was done using three measures suggested by \citet{RN180}. These were the Pearson's correlation coefficient ($r$), the squared Pearson's correlation coefficient or determination coefficient (R$^2$), the modelling efficiency coefficient (MEF; eq. 4) and the reliability index (RI; eq. 5). The model efficiency coefficient and the reliability index are described as: \par

\begin{equation}%
MEF =  \frac{\left(\sum_{i=1}^{n}(O_i - \bar{O})^2 - \sum_{i=1}^{n}(P_i - O_i)^2\right)}  {\sum_{i=1}^{n}(O_i - \bar{O})^2}
\end{equation}
\par
\bigskip

And:
\par

\begin{equation}%
RI = exp \sqrt{\frac{1}{n} \sum_{i=1}^{n} \left(log\frac{O_i}{P_i}\right)^2}
\end{equation}
\par

\bigskip
Here, $n$ is the number of observations, $O_i$ is the $ith$ of observation, $\bar{O}$ is the average of the observations and $P_i$ is the $ith$ prediction.\par
The correlation coefficient, $r$, can range from -1 to 1. An $r$ value of 0 indicates no correlation, whereas values of 1 indicate that the simulated and reference values variate with an equal trend. Negative $r$ values indicate inverse variation between simulated and reference values. 
MEF can vary between $-\infty$ and 1. Negative values of MEF indicate that the model has a worse fit to the data than an average through the time series; $MEF=0$ indicates that the model has an equal fit to an average through the time series; and $MEF=1$ indicates that the model perfectly fits to the time series.
Models with negative MEF can still be useful, if $r$ is positive. Negative MEF values could be due to differences in magnitude between simulated and observed values, despite the simulations following the same trend as the observations.
Finally, RI is a measure of the difference in magnitude between simulated and reference values. It varies between 1 to $+\infty$ and a value of 1.5 would indicate that the model simulates the reference values with a 50\% difference in magnitude.\par

\section{Results and Discussion}

\subsection{Mass-balance model}
The estimated biomass for capelin in 1984 was 4 681 408 tonnes, $2.4$ times higher than MFRI estimates for that year \citep{RN148}. These results are consistent with a previous study by \citet{RN102}, in which the authors hypothesised that capelin biomass was being grossly underestimated in the assessments and should be approximately doubled in order to explain the amount of capelin that is being consumed by predators.\par
The combined predation mortality of cod on capelin accounted for $31\%$ of $M$ of capelin ($M=0.311$; total $M=1.014$). Blue whiting had the second highest effect on capelin $M$, with its predation mortality on capelin accounting for $22\%$ of $M$ ($M=0.224$).\par

\subsection{Dynamic model}
Overall, the Ecosim model performed in a satisfactory manner when replicating previously known dynamics (Table 6 and Figures 2 and 3).
The anomaly search using 10 spline points resulted in the fit with the lowest AIC \tablename{ 5}. The anomaly search on the final model was thus done using 10 spline points.
When fitting the model to the time series of biomass and landings by estimating the vulnerability parameters, the original fit ($SSQ=84.80$) was improved by 4\% ($SSQ=81.56$). After the addition of temperature as a forcing function of PP to the time series of biomass, $F$ and landings, the original fit was improved by 16\% ($SSQ=71.04$). Considering these results, temperature was included in the final model as a forcing function of PP.\par

\begin{table}[H]
	\begin{center}
		\caption{Summary of SSQ and AIC using different number of spline points.}\par
		\footnotesize
		\begin{filecontents*}{Summary.csv}
Spline points,SSQ,AIC
0,75.52,-880.4
5,75.01,-944.5
10,71.04,-961.4
15,72.17,-942.4
20,68.53,-957.5
\end{filecontents*}
\csvautobooktabular{Summary.csv}\par
		\normalsize
	\end{center}
\end{table}

The best fit was achieved when annual temperature was included in the model, suggesting that fisheries alone cannot explain the Icelandic marine ecosystem.
Temperature is considered to be one of the main determining factors in marine fish distribution \citep{RN210, RN211}. Cod has been shown to change its habitat use in the southern North sea according to surface temperature \citep{RN209}. Considering the fundamental role temperature plays in fish ecology, it is not surprising that the addition of temperature to the model resulted in a better fit.

\par

\begin{table}[H]
	\begin{center}
		\caption{Skill assessment per group, where r is the Pearson's correlation coefficient, R$^2$ is the determination coefficient, MEF is the modelling efficiency and RI is the reliability index.
	    Reference values for a good fit: $0.7\leq r\leq 1$; $0.5\leq R^2\leq 1$; $0<MEF\leq 1$; $1\leq RI \leq 1.5$.}\par
		\footnotesize
		\begin{filecontents*}{SA.csv}
Functional group,, Biomass,,,,Catches,,
,r,R$^2$,MEF,RI,r,R$^2$,MEF,RI
Cod 4+,0.605,0.366,-1.079,1.291,0.810,0.656,0.172,1.291
Saithe 4+,0.750,0.563,0.599,1.242,0.844,0.712,0.479,1.242
Haddock 3+,0.768,0.590,0.635,1.241,0.623,0.388,0.471,1.241
Herring 4+,0.323,0.104,-0.223,1.398,0.577,0.333,0.047,1.398
Redfish 8+,0.567,0.321,0.147,1.230,0.796,0.634,0.279,1.230
Greenland halibut 6+,,,,,0.188,0.035,-0.354,1.909
Capelin,,,,,0.763,0.583,-0.419,2.006
Blue whiting,0.535,0.286,-0.012,1.909,0.929,0.863,0.424,1.755
Mackerel,,,,,1.000,1.000,1.000,1.000
Flatfish,,,,,0.778,0.605,0.391,1.238
Other codfish,,,,,0.754,0.568,0.506,1.251
Comercial demmersal,,,,,0.687,0.472,-0.444,1.318
\end{filecontents*}
\csvautobooktabular{SA.csv}\par
		\normalsize
	\end{center}	
\end{table}

The skill assessment had overall satisfactory results ($0.7\leq r\leq 1$; $0.5\leq R^2\leq 1$; $0<MEF\leq 1$;  $1\leq RI \leq 1.5$; \tablename{ 6}). There was a general tendency from the model to replicate the landings better than the biomass. With the exception of group haddock 3+, fitted biomass had higher R$^2$ values than fitted landings (\tablename{ 6}). These results are unsurprising, as the landings were actual observations, while the biomass used were previous estimates by the MFRI \citep{RN148}.\par
There were three groups (cod 4+, herring 4+ and blue whiting) for which the model had a worse fit for biomass than a straight line through the average of the reference values ($MEF<0$; \tablename{ 6}). The model performed poorly when predicting the landings of Greenland halibut 6+, capelin and commercial demersal fish as well ($MEF<0$; \tablename{ 6}). The MEF is sensitive to the difference between the simulated and the reference values (Eq. 4) and can be negative even when the R$^2$ is within reasonably good range ($0.5\leq R^2\leq 1$), as was the case of capelin landings (\tablename{ 6}). The model underestimated capelin landings with a difference in magnitude of approximately 100\% ($RI=2.006$; \tablename{ 6}), which could account for the low MEF value.\par
The model was calibrated using biomass estimates from single-species assessments, as well as $F$ calculated using those estimates. This method was chosen, as there were no biomass estimates from multi-species assessments available and because of how Ecopath internally computes $F$ as $Y/B$. The use of biomass estimated by single species assessments in a multi-species context could have contributed to the poor skill assessments results observed for herring 4+ and redfish 8+ biomass, as well as for Greenland halibut 6+ landings.\par

\begin{figure}[H]
	\includegraphics[scale=0.5]{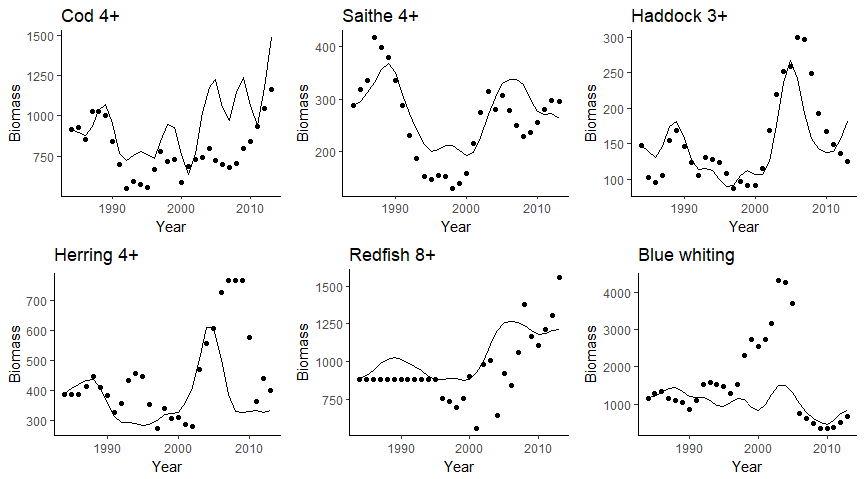}\par
	\caption{Biomass estimates by year for all groups and reference biomass time-series. Estimates by Ecosim are represented by a solid line and time-series reference data is represented by dots. Biomass is thousands of tonnes. Notice the different scales.}\par
\end{figure}

With the exception of mackerel, all groups had $RI>1.2$, which indicates that the simulated biomass and landings varied at least 20\% in magnitude in relation to the reference biomass and landings (\tablename{ 6}). However, none of these differences in magnitude were constant through the simulations, indicating that there were no systematic errors (Figures 2 and 3).\par
The skill assessment of mackerel caches shows optimal results for all measures (\tablename{ 6}). These can be explained by the use of forced biomass for this group. Using forced biomass for mackerel in EwE can have consequences for the modelled trophic dynamics, as it could lead to over-fitting. Over-fitting the model can, in turn, lead to unnatural simulations of the groups with direct trophic links to this group. However, the use of fisheries and temperature was not enough to satisfactorily simulate the invasion by mackerel. \par

\begin{figure}[H]
 \includegraphics[scale=0.5]{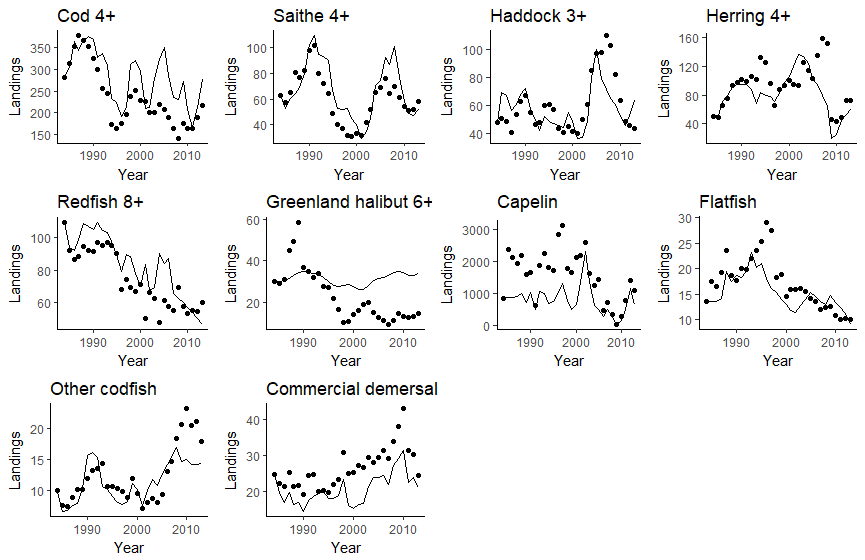}\par
 \caption{Landings estimates by year for all groups and reference landings time-series. Estimates by Ecosim are represented by a solid line and time-series reference data is represented by dots. Catches are in thousands of tonnes. Notice the different scales.}\par
\end{figure}

\subsubsection{Trophic flows}

Ecosim uses a vulnerability scale of 1 to $+\infty$, where $v=2$ represents mixed-trophic control, $v<2$ represents bottom-up control and $v>2$, represents top-down control \citep{RN97}. The higher the vulnerability, the stronger top-down effect a predator has on its prey, this being relevant for values up to $v=100$.\par 
Of the groups with vulnerability parameters freed from default, only cod 4+ had $v<2$ (\tablename{ 7}), suggesting that this group is affected by the abundance and/or availability of their prey, as contrast to acting as a controlling force over them.
The groups capelin, blue whiting and commercial demersal had $v=1.979$, $v=2.156$ and $v=2.000$, respectively (\tablename{ 7}), which suggests that these groups have mixed trophic control. The remaining groups had $v>100$, suggesting top-down control from these groups over their prey (\tablename{ 7}). Only high TL predators ($TL>3$; Figure 4) had their vulnerability parameters freed from default (\tablename{ 7}), as these were the only groups with time-series of biomass and/or landings available, and these usually are linked to their prey through top-down control. As such, it is not possible to speculate about which type of trophic control dominates the Icelandic marine ecosystem from this model.\par

\begin{figure}[H]
	\begin{center}
		\includegraphics[scale=0.32]{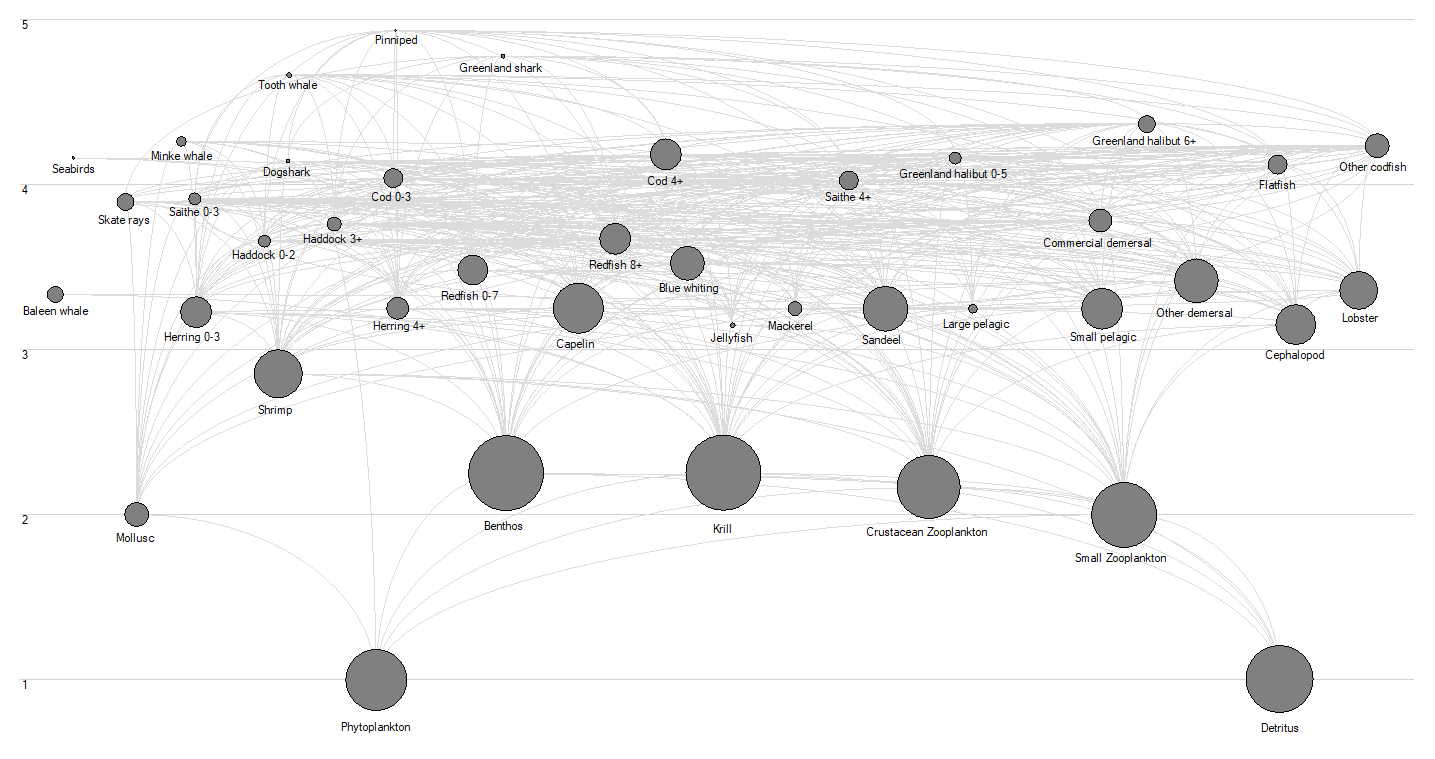}\par
		\caption{Flow diagram for the Ecopath model of Icelandic waters in 1984 showing the distribution of the functional groups by trophic level. Larger nodes represent bigger stock sizes. The trophic level scale can be found on the left of the diagram.}\par
	\end{center}
\end{figure}

The vulnerability search was done by predator. When freeing vulnerabilities from default in this manner, it is assumed that how vulnerable a prey is to their predator depends mostly on the predator's foraging behaviour \citep{RN17}. 
Conducting the vulnerability search by predator will thus lead to one vulnerability value for every predator-prey interaction of a given predator \citep{RN17}. 
Assuming that a predator's foraging behaviour remains unchanged independently of the prey sought might not always depict reality. 
An example of this is cod, which was estimated to be affected by bottom-up control ($v<2$; \tablename{ 7}) by this model. 
However, a previous study has shown that cod has top-down control over shrimp, but is affected by bottom-up control by capelin \citep{RN178}. 
The vulnerability search method used prevents individual predator/prey pairs from having a unique vulnerability value. 
The vulnerability search was done in this manner as it was the only option available to restrict it to the groups with time series.\par

\begin{table}[H]
	\begin{center}
		\caption{Estimated vulnerabilities ($v$)for predator.}\par
		\footnotesize
		\begin{filecontents*}{Vulnerabilities.csv}
Predator,$v$,Functional group,$v$
Cod 4+,1.186,Capelin,1.979
Saithe 4+,$>100$,Blue whiting,2.156
Haddock 3+,$>100$,Mackerel,$>100$
Herring 4+,$>100$,Flatfish,$>100$
Redfish 8+,$>100$,Other codfish,$>100$
Greenland halibut 6+,$>100$,Commercial demersal,2.000
\end{filecontents*}
\csvautobooktabular{Vulnerabilities.csv}\par
		\normalsize
	\end{center}
\end{table}

\subsection{Comparisson with other models}

There were two previously published Ecopath models describing the Icelandic marine ecosystem in 1950 and 1997 \citep[Buchary, E.A.; Mendy, A.N and Buchary, E.A., respectively, in][]{RN208,RN171}. 
The model for 1950 included an Ecosim simulation, ran from 1950 to 1997.\par
In the 1950 study, the author constructed the model to be comparable to the Ecopath model for 1997 \citep[Buchary, E.A. in][]{RN208}.
The 1950 model Ecopath was an attempt to reconstruct the ecosystem for that year \citep[Buchary, E.A. in][]{RN208, RN171}, by estimating the biomass for most functional groups. The author then used a time series of F and reference biomass (calculated as Y/F) of cod and herring to fit the model \citep[Buchary, E.A. in][]{RN208}. 
No vulnerability search was done in this study and all vulnerabilities were set to 0.3 \citep[Buchary, E.A. in][]{RN208}.
There were differences between the input biomass and landings of the 1997 model \citep[Mendy, A.N and Buchary, E.A. in][]{RN208,RN171} and the simulations by the present model for the same year for most groups. It is reasonable to think that these discrepancies were due to different modelling assumptions such as modelled area and species density, as the differences are also found between the input data of the older model and the data sources for the current model.\par
Both the 1950 and 1997 models used 25 functional groups each \citep[Buchary, E.A.; Mendy, A.N and Buchary, E.A., respectively, in][]{RN208,RN171} in contrast to the 35 functional groups of the here presented model. 
The current model was more complete than the 1950 and 1997 models, as it included more detailed functional groups, a more complete time series than the one used in the 1950 model and because of it being fitted to the time series through vulnerability and anomaly searches. Furthermore, a skill assessment was carried out in the current model, which was not the case for the previous models.
\par

\subsection{Final remarks}
It was estimated by this model that commercial fishes like haddock, redfish, Greenland halibut, flatfish and other codfish have top-down control over their prey. These being predators implies that their more or less heavy removal from the system could have serious implications for the ecosystem dynamics and should be taken into account by fisheries management authorities. Furthermore, mid trophic level species like herring and mackerel were also estimated to have top-down control over their prey. These species have a direct trophic link to lower trophic level organisms and their exploitation should surely take this into account.\par
The use of EE to estimate biomass in Ecopath, might raise concerns, as it is inversely related to biomass. An alternative to using EE could be using species density from other EwE models in neighbouring areas. However, using this method would imply assuming that either the marine ecosystem in Iceland would have the same physical, hydrological and biological dynamics as the neighbouring areas or that species density is independent of these. This would be an assumption with potentially serious consequences, as the distribution of species in marine systems is thought to be linked to biotic and abiotic factors \citep{RN211}.\par
This model provides insight on the marine trophic web in Iceland and it could be used as a tool to investigate how different fishing policies could impact the trophic web dynamics in Icelandic waters.\par

\section{Acknowledgements}
This study has received funding from the European Union’s Seventh Framework Programme for research, technological development and demonstration under grant agreement no. 613571 for the project MareFrame and from the European Commission’s Horizon 2020 Research and Innovation Programme under Grant Agreement No. 634495 for the project Science, Technology, and Society Initiative to minimize Unwanted Catches in European Fisheries (Minouw), as well as from the Icelandic Research Fund (Rannis, No. 152039051).\par
Furthermore, the authors would like to acknowledge the MFRI for providing the data that made the work presented here possible, Maciej T. Tomczak, in providing useful comments on an early version of the paper and Villy Christiansen for giving helpful input on the time series fitting.

\section*{References}

\bibliography{EwE_Paper_JR}

\pagebreak
\section*{Appendix A}

\footnotesize
\tablename{ A.1:} Description to the species level of the functional groups used in the EwE model for Icelandic waters.\par
\csvautobooklongtable{FGComposition.csv}
\normalsize

\pagebreak

\footnotesize
\tablename{ A.2:} Input data for the balanced Ecopath model for Icelandic waters. Biomass and landings are given in thousands of tonnes.\par
\csvautobooklongtable{Input.csv}
\normalsize

\pagebreak

\footnotesize
\tablename{ A.3:} Output for the Ecopath model for Icelandic waters. Biomass is given in thousands of tonnes.\par
\csvautobooklongtable{Output.csv}
\normalsize

\end{document}